\documentclass[review]{elsarticle}

\usepackage{lineno,hyperref}
\usepackage{amsfonts}
\usepackage{amsmath,amssymb}
\usepackage{xcolor}
\modulolinenumbers[5]

\journal{American Journal of Physics}

\begin{document}

\begin{frontmatter}

\title{{Still learning about space dimensionality: from} the description of hydrogen atom by a generalized wave equation
{for dimensions D$\geq$3}}


\author[mymainaddress,mysecondaryaddress]{Francisco Caruso}\corref{mycorrespondingauthor}
\cortext[mycorrespondingauthor]{Corresponding author}
\ead{francisco.caruso@gmail.com or caruso@cbpf.br}

\author[mysecondaryaddress]{Vitor Oguri}
\author[mysecondaryaddress]{Felipe Silveira}

\address[mymainaddress]{Centro Brasileiro de Pesquisas F\'{\i}sicas -- Rua Dr.~Xavier Sigaud, 150, 22290-180, Urca, Rio de Janeiro, RJ, Brazil}
\address[mysecondaryaddress]{Instituto de F\'{\i}sica Armando Dias Tavares, Universidade do Estado do Rio de Janeiro -- Rua S\~ao Francisco Xavier, 524, 20550-900, Maracan\~a, Rio de Janeiro, RJ, Brazil}

\begin{abstract}
The hydrogen atom is supposed to be described by a generalization of Schr\"{o}dinger equation, in which the Hamiltonian depends on an iterated Laplacian and a Coulomb-like potential $r^{-\beta}$. Starting from previously obtained solutions for this equation using the $1/N$ expansion method, it is shown that new light can be shed on the problem of understanding the dimensionality of the world as proposed by Paul Ehrenfest. A surprisingly new result is obtained. Indeed, for the first time, we can understand that not only the sign of energy but also the value of the ground state energy of hydrogen atom is related to the threefold nature of space.

\end{abstract}


\end{frontmatter}

\section{Introduction}
\label{intro}

The possibility that extra spatial dimensions can play important role in Physics is not new. It can be traced back to the pioneer works of Kaluza~\cite{Kaluza} and Klein~\cite{Klein}.
Following the general unification idea of Kaluza-Klein~\cite{Appel}, several higher-dimensional theories were developed, like String Theory and Supersymmetry~\cite{Schwarz,Dine}, based on theoretical ideas that go beyond the Standard Model of Particle Physics and show promise for unifying all forces.

Indeed, the introduction of extra dimensions in the physics of fundamental interactions has enabled a remarkable progress in two major contemporary programs: the quantization of gravity and the unification of the force fields of Nature, for which the mechanisms of reduction and dimensional compactification are of utmost importance~\cite{Helayel}.

Around 1980, String Theory proposed to enlarge the number of space dimensions, in this instance as a requirement for describing a consistent theory of Quantum Gravity. The extra dimensions were supposed to be compactified at a scale close to the Planck scale, and thus not testable experimentally in the near future~\cite{PDG}.

However, in 1998, a new approach was proposed~\cite{Arkani}, where it was shown that the weakness of gravity could be explained by postulating two or more extra dimensions in which only gravity could propagate. In particular, it was argued that the hierarchy problem of the Standard Model of Elementary Particles could be solved within this postulate. It is important to stress that, in this case, the size of these extra dimensions should vary on a scale accessible to experimentation (ranging between roughly $0.001$ and $1.97 \times 10^{-13}$ meters), leading, therefore, to observable consequences in current and future experiments at Large Hadron Collider (LHC/CERN)~\cite{Agashe,Black}. So, for the first time, it was realized that extra dimensions could also have an important impact on an open problem in Particle Physics at energies $\simeq 1$~TeV. Other applications are reviewed in Refs.~\cite{Agashe,Lorenzana}.

\textit{In summa}, all these developments emphasize the idea that the existence of extra dimensions could be one of the most attractive possibilities for Physics beyond the Standard Model, and that such idea may be experimentally tested.

In Section~\ref{chal}, some works, aimed at understanding space dimensionality, are briefly reviewed, and their epistemological limitations are pointed out. In Section~\ref{Nexp}, the $1/N$ expansion method to a form that will be used in Section~\ref{general} to compute the energy levels of the hydrogen atom ground states for a generalized equation in a $D$-dimensional flat space is revised. In this Section~\ref{general}, a more general way of describing the hydrogen atom is presented in $D$ dimensions, which to a certain extent, {avoid the epistemological criticisms of} Section~\ref{chal}. Results are presented and discussed in Section~\ref{sec:4}, and concluding remarks are made in Section~\ref{sec:5}.


\section{Epistemological challenge}
\label{chal}

All that was said in the Introduction brings back an old question:
{How are the} fundamental laws of Physics and space-time dimensionality entangled? The genesis of this kind of investigation can be traced back to the doctoral thesis of Immanuel Kant \cite{Kant}. The role played by space dimensionality in determining the form of various physical laws and constants of nature was reviewed in Ref.~\cite{Barrow}. {For more insights on the subject, see also \cite{Bianchi}.}

A systematic scientific investigation of this general question begins with the seminal contributions of Paul Ehrenfest \cite{Ehrenfest1,Ehrenfest2}. His idea was to identify particular aspects of a physical system or phenomenon, called by him ``singular aspects'', which could be used to distinguish the Physics in three-dimensional space from that in $D$-dimensions. To carry out this project, he started from postulating that the form of a differential equation, -- which usually describes a physical phenomenon in a three-dimensional space --, is still valid for an arbitrary number of dimensions. As an example, Ehrenfest assumed the motion of a planet under a central force associated with the Newtonian gravitational potential to be still described by the Laplace-Poisson equation, keeping the same power of the Laplacian operator $\Delta$ and altering the number of spatial coordinates from 3 to $D$. Based on the general mathematical solution of such equation, he imposed that they should satisfy the postulate of the stability of orbital motion under central forces to get at the proper number of dimensions. It is clear that, in this case, one cannot claim to have demonstrated that $D=3$ for an obvious reason: The Poisson equation for the Newtonian potential is, by construction, an equation that effectively describes stable orbits in three dimensions. Thus, Ehrenfest's result was already expected and, actually, it has to be seen almost as a consistency check of the theoretical description of stable planetary motion. For an epistemological critique of Ehrenfest's work see \cite{Moreira}.

Using a semi-classical approach, Ehrenfest, basing his argument on the Niels Bohr quantization of circular atomic orbits generated by a Coulomb-like potential -- formally analogous to the Newtonian one -- also showed that there is no bound state for the hydrogen atom when $D\geq 5$.

In 1963, Tangherlini was the first to formally treat the problem of the hydrogen atom from the point of view of Schr\"{o}dinger equation \cite{Tangherlini}. This article inspired several others that will be cited throughout the text. However, it is important to mention that even here the operator of the kinetic part of the generalized Schr\"{o}dinger's Hamiltonian is assumed to be still $\Delta$ in $D$-dimensions.

The postulate that the form of a given equation -- obtained without questioning space dimensionality, as has always been the case -- is still valid for arbitrary $D$ (up to the potential energy term that can change with dimensionality, as it is the case for Newtonian and Coulomb-like potentials) is the Achilles' heel of any scientific discussion of space dimensionality. This is quite unavoidable. The only thing we can do is to change perspective and ask ourselves, as Bollini and Giambiagi did \cite{Bilbao}: -- Are some physical theories related to a specific number of dimensions? This means, in a certain sense, that we must go further, so far the generalization of a mathematical law is concerned, in order to bypass the aforementioned epistemological critique. The original equation has to be generalized into a class of plausible hypothetical equations (for which it is just a particular case) that could describe our physical system in $D$ dimensions, instead of simply by enlarging the number of dimensions, while keeping the same original operator structure of the original law.

Inspired by this alternative point of view and with this strategy in mind, the generalization of a particular equation must foresee the appearance of extra integer powers ($n$) in Hamiltonian operators (for example, $\Delta \rightarrow \Delta^n$), as well as the substitution of the potential by the generalized one and, finally, this new equation will be assumed to depend on $D$ coordinates. Doing so, and supposing we are able to analytically solve this new equation, we can try to identify some special features associated with the set of parameters $(D,n)$ and not only to the dimensionality $D$. In particular, we can investigate whether what could be called a ``singular aspect'' of Physics in $D=3$ would not be a unique feature of a particular combination of both parameters $D$ and $n$. Following these ideas, let us try to construct here one example based on the hydrogen atom.

\section{$1/N$ expansion method in atomic physics}\label{Nexp}
{
Initially proposed by Gerard 't Hooft in 1974 \cite{Hooft, Hooft2}, and developed by several authors~\cite{Hagen, Bjerrum, Chatterjee, Imbo, Kais, Mlodinow, Witten} the $1/N$ expansion method is also known as large $N$ expansion. Originally, it had the purpose of overcoming certain mathematical difficulties found in QCD but were soon applied to Atomic Physics. To better understand the motivation behind this expansion, let's start by looking at the usual Hamiltonian of the hydrogen atom in $D=3$:}

\begin{equation}
 H = \frac{p^2}{2m} - \frac{e^2}{r},
\end{equation}

{\noindent we could treat the potential energy as a perturbation, if we chose to take $e^2$ as a very small value. But this choice would not make sense physically, as the constant $e$ has a specific dimension and cannot be simply changed; the value of $e$ depends only on the chosen units.}

{When we make the change of variables $r = \rho/me^2$ and $p = P m e^2$, we can rewrite the Hamiltonian as:}
\begin{equation}
H = me^4 \left(\frac{P^2}{2} - \frac{1}{\rho}\right)
\end{equation}
Choosing a convenient set of units ($\hbar=m=e=1$) the above Hamiltonian can be reduced to the form

\begin{equation}
\widehat{H} = \left(\frac{P^2}{2} - \frac{1}{\rho}\right).
\end{equation}
{We observe that now this Hamiltonian does not have any free parameters, and therefore, we cannot do a perturbative expansion. Of course, we know that this kind of system can be easily solved, but, suppose that wasn't the case, and that even a numerical solution was impractical. What could be done?}

{We need to choose a parameter, which is usually given as fixed and known, and treat it as a free variable parameter. According to 't Hooft, such a parameter is the dimensionality of space. Once we define space as no longer having a fixed number of dimensions, we can observe that atomic physics becomes simpler when we take $N\rightarrow \infty$ and that it can be solved for large $N$ by an expansion in $1/N$.}

{In the next section we will undertake the step-by-step expansion for the case of a wave equation that describes the hydrogen atom.}

\section{Generalized hydrogen atom in a $D$-dimensional flat space}\label{general}

{No one doubts that if space has a total number of dimensions $D$, this number should be greater than 3. This is why we were not concerned with either the case $D=1$ or $D=2$.}

We start from the hypothesis that a hydrogen atom is described by the wave equation
 \begin{equation}\label{wave-eq}
(-1)^n \Delta^n \psi - \frac{\alpha}{r^\beta}\psi = E \psi
\end{equation}

\noindent {where $\Delta$ is the $D$-dimensional Laplacian operator, $n$ is an integer, $\alpha$ and $\beta$ are parameters and $E$ is an eigenvalue. Since the Coulomb-like potential $\sim r^{-\beta}$ corresponds to a long range force such that $V(r) \rightarrow 0$ when $r \rightarrow \infty$, we must have $\beta\geq 0$, where the particular case $\beta=0$ is associated to a $\ln(r)$ dependence. For simplicity, we will restrict ourselves to spherically symmetric solutions for the $s$-wave state. The Laplacian operator can be written as}

$$\Delta = \frac{d^2}{dr^2} + \frac{D-1}{r}\frac{d}{dr}$$

{The derivation of equation~(\ref{wave-eq}) can be found in the references~\cite{Bjerrum, Chatterjee, Dong}, for $n=1$. Therefore, didactically, let us first consider the case $n=1$. In this case, we have}

 \begin{equation}\label{wave-eq2}
- \left(\frac{d^2}{dr^2} + \frac{D-1}{r}\frac{d}{dr}\right)\psi - \frac{\alpha}{r^\beta}\psi = E \psi
\end{equation}

Making the transformations $\psi = r^a \phi$ with $a = \frac{1}{2} (1-D)$ we can eliminate the first order derivative of $\phi$ using
$$\frac{d}{dr} (r^a \phi) = a r^{a-1} \phi + r^a \frac{d\phi}{dr} $$
\noindent yielding
$$\left(\frac{d^2}{dr^2} + \frac{D-1}{r}\frac{d}{dr}\right) r^a \phi = r^a \frac{d^2\phi}{dr^2} - a(a+1) r^{a-2} \phi $$

\noindent {Then, equation~(\ref{wave-eq2}) becomes}
 \begin{equation}\label{wave-eq3}
- \left(r^a \frac{d^2}{dr^2} - a(a+1) r^{a-2}\right)\phi - \frac{\alpha}{r^\beta}r^a \phi = E r^a \phi
\end{equation}

\noindent and, replacing $r = D^\tau R$ with $\tau = \frac{2}{2-\beta}$ (for $n=1$),
 \begin{equation}\label{wave-eq4}
-\left(D^\tau R\right)^a D^{-2\tau} \frac{d^2}{dR^2}\phi + a(a+1) \left(D^\tau R\right)^{a-2}\phi - \frac{\alpha}{\left(D^\tau R\right)^\beta}\left(D^\tau R\right)^a \phi = E \left(D^\tau R\right)^a \phi
\end{equation}

\noindent which, by eliminating the term $\left(D^\tau R\right)^a$, can be simplified as

 \begin{equation}\label{wave-eq5}
- D^{-2\tau} \frac{d^2}{dR^2}\phi + \frac{a(a+1)}{D^{2\tau} R^2}\phi - \frac{\alpha}{D^{\beta\tau} R^\beta} \phi = E \phi
\end{equation}

\noindent or, multiplying by $\frac{D^{2\tau}}{D^2}$,
 \begin{equation}\label{wave-eq6}
- \frac{1}{D^2} \frac{d^2}{dR^2}\phi + \frac{a(a+1)}{D^{2} R^2}\phi - \frac{\alpha}{ R^\beta} \phi = E \frac{D^{2\tau}}{D^2} \phi
\end{equation}

\noindent {If we take the limit for large $D$, the kinetic part of equation~(\ref{wave-eq6}) is suppressed giving rise to}

 \begin{equation}\label{wave-eq7}
\left(\frac{1}{4 R^2} - \frac{\alpha}{R^\beta}\right) \phi \simeq E D^{2\tau - 2} \phi
\end{equation}

\noindent In this case, the particle is at the minimum of the effective potential, this means that the ground state energy is the absolute minimum and the excitation energies could be calculated by quadratic approximations near it minimum. Rewriting equation~(\ref{wave-eq7}) as $V_{\text{eff}} \phi = E\phi$, we obtain
$$V_{\text{eff}} = \left(\frac{1}{4 R^2} - \frac{\alpha}{R^\beta}\right) \frac{1}{D^{2\tau - 2}}$$

\noindent which has a minimum at
$$R_0 = (2 \alpha \beta)^{\frac{1}{\beta - 2}}$$

This same process can be repeated for other values of $n$, and for the generalized case we must use the following Laplacian~\cite{Gelfand}
$$(-1)^n \Delta^n r^a = 2^{2n}(\zeta +1)\ldots (\zeta+n)(\zeta+\frac{1}{2}D)\ldots \times (\zeta + \frac{1}{2}D +n -1)r^{a-2n}$$

\noindent {with $\zeta = \frac{1}{4} (1-D) -n = \frac{1}{2} a-n$. Following the same steps we did in the case $n=1$, we obtain the effective potential}
\begin{equation}\label{veff}
V_{\text{eff}} = \left(\frac{\Lambda}{D^{2n} R^{2n}} - \frac{\alpha}{R^\beta}\right)\frac{1}{D^{2\tau -2}}
\end{equation}

\noindent {where, now $\tau=2n/(2n-\beta)$, and $\Lambda$ is a number dependent on both $n$ and $D$, which, for a large value of $D$, has the behavior of $\Lambda(n,D) = \left(\frac{1}{2} D\right)^{2n}$.
Taking the first derivative of equation~(\ref{veff}) we obtain}
$$\frac{d V_{\text{eff}}}{dR} = D^{-\beta \tau} \left(\alpha \beta R^{-\beta-1} - 2\Lambda n D^{-2n} R^{-2n-1}\right)$$

\noindent {By equating it to zero, we determine the distance $R_0$ which minimizes the energy:}
$$R_0^{2n-\beta} = \left(\frac{2\Lambda n}{D^{2n}\alpha \beta}\right)$$

\noindent {which, for large $D$, values can be written as}
$$R_0 = \left(\frac{2 n}{2^{2n}\alpha \beta}\right)^{\frac{1}{2n-\beta}}$$

Substituting the $R_0$ value in the effective potential, we get finally an approximate expression for the ground state energy~\cite{Giambiagi}:


\begin{equation}\label{E0}
E_0 = - \, \frac{\alpha}{D^{2n\beta/(2n-\beta)}\, \displaystyle\left(\frac{2n}{2^{2n}\alpha\beta}\right)^{\beta/(2n-\beta)}}\, \frac{2n-\beta}{2n}
\end{equation}

Here, atomic units ($\hbar = m = e = k = 1$) are adopted in which energies are expressed in hartree. With this convenient choice, the above formula for $E_0$ depends only on three independent integer parameters $D$, $n$ and $\beta$, as considered in Ref.~\cite{Giambiagi}.

However, $\beta$ should depend on $D$. Actually, it is well known that the potential should behave as $1/r^{D-2}$~\cite{Ehrenfest1,Ehrenfest2,Tangherlini,Mostepanenko,Supplee,Shaqqor,Caruso,Dirac}, for $D\geq 3$, and as $\ln r$, for $D=2$, if it is the solution of the $D$-dimensional equation $\Delta \varphi (r)= -4\pi\delta(r)$. This choice ensures, at least at classical level, the electric charge ($e$) conservation~\cite{Supplee} which follows from the integral form of Gauss law. However, this is not the most general choice for the potential, as we will see. Indeed, if we consider, as in equation~(\ref{wave-eq}), that the operator $\Delta$ should also be replaced by $\Delta^\lambda$, with $\lambda$ being an integer, another relation between $\beta$, $D$ and $\lambda$ will result. The most general Poisson equation, in this case ($e=1$), is of the form
\begin{equation}\label{general-poisson}
(\Delta)^\lambda G(r) = -4\pi \delta(r)
\end{equation}
So, $\beta$ can still depend on $\lambda$.

Before we proceed to get numerical values for $E_0$, we must fix the dependence of both $\alpha$ and $\beta$ parameters of the generalized Coulomb potential, $V(r)$, on space dimensionality $D$ and on the power ($\lambda$) of the iterated Laplacian which appears in the radial part of the generalized Poisson equation~(\ref{general-poisson}) having the potential $-\alpha\,r^{-\beta}$ as solution.

Equation~(\ref{general-poisson}) was studied in detail in Ref.~\cite{Bollini93} where the Green function $G(r)$ -- which is exactly the Coulomb-like potential for a point-like unit charge ($e=1)$ -- was calculated, resulting
in the following expression for the generalized potential $V(r)$ to be used in equation~(\ref{wave-eq}):
\begin{equation}\label{pot-V-general}
V(r) = \frac{(-1)^{\lambda+1}\, \Gamma(D/2-\lambda)}{4^{\lambda-1} \pi^{D/2-1}\, \Gamma(\lambda)}\, \frac{1}{r^{D-2\lambda}} \equiv \frac{\alpha(D,\lambda)}{r^{D-2\lambda}} \equiv \frac{\alpha}{r^\beta}
\end{equation}

It is important to stress that, in order to have a power law of the type $r^{-\beta}$ for the potential, the parameter $\beta$ should satisfy the condition $\beta = D-2\lambda \geq 0$. Thus, in principle, the ground state energy depends on just two parameters, \textit{i.e.}, $E_0 = E_0(D,\lambda)$.

For $\lambda =1$ and arbitrary $D$, we get from equation~(\ref{pot-V-general}) the well known result \cite{Supplee,Caruso}
\begin{equation}\label{V-m1}
V(r) = \frac{2\Gamma(D/2)}{\pi^{(D/2-1)}}\, \frac{1}{(D-2)r^{D-2}} \equiv \frac{\alpha(D,1)}{r^{D-2}}
\end{equation}
and, for $\lambda=1$ and $D=3$, there follows the usual three-dimensional potential
\begin{equation}
V(r) = \frac{1}{r}
\end{equation}

From a simple inspection of equation (\ref{pot-V-general}), we conclude that whenever $\lambda$ is an even integer ($\lambda=2\ell,\, \ell=1,2,3,\cdots$) the nature of the potential changes from attractive to repulsive. Therefore, there is no bound states for \textit{even} values of $\lambda$.

Nonetheless the above relation between $\beta$, $D$ and $\lambda$ is not an unique choice. In Ref.~\cite{Burg}, for example, the authors made one more assumption by fixing $\lambda=(D-1)/2$ in order to assure that the potential has always the form $r^{-1}$, a result demonstrated in Ref.~\cite{Bilbao}. The price to pay is very high, since the Gauss law is no longer valid, leading to the necessity of modifying Maxwell equations in higher dimensions. In any case, we should be aware of Hermann Weyl's classical result \cite{Weyl1,Weyl2,Weyl3} which tell us that Maxwell equations lose scale invariance if space-time dimension is different from $3+1$.

{For $D=3$, we know that $\lambda=1$ and $n=1$.} Inspired in what happens in three dimensions, we will assume throughout the paper the validity of Gauss law {for any $D$. The simplest choice consistent with these facts is $\lambda=n$.} So, the ground state energy, associated with equation (\ref{wave-eq}) for a potential given by equation (\ref{pot-V-general}), is obtained by substituting $\beta = D-2n$ in equation~(\ref{E0}), yielding
\begin{equation}\label{E0-final}
E_0 (D,n) = - \, \displaystyle\left[\frac{n (D/2)^{2n}}{(D/2-n)}\right]^{\frac{(D/2-n)}{(D/2-2n)}}\times \alpha(D,n)^{-n/(D/2-2n)}\, \times  \frac{2n-D/2}{n}
\end{equation}

\section{Results and discussions}
\label{sec:4}

Thus, in principle, $E_0 < 0$ if $4n-D > 0 \, \,\Rightarrow \, \,D < 4n$. (For $D=4n$, $E_0$ diverges.) In such cases, the potential has a ground state and it is said to be non-singular. Whenever $D>4n$ the extremum of the potential is a maximum. In this case the effective potential has no minimum and it is said to be singular. Therefore, the existence or not of a negative energy state as a solution of equation (\ref{wave-eq}) depends on two integer numbers: the power of the Laplacian $(n)$ and space dimensionality $(D)$. However, in the case $m=n$, we must also take into account the constraint $\beta = D-2n > 0$ (due to the long range nature of the potential), which means that
\begin{equation}\label{condition}
2n<D<4n
\end{equation}
So, for each power $n$ of the Laplacian there is a minimum value of space dimensionality, given by $D_{\mbox{\tiny{min}}} = 2n+1$, for which a bound state {does} exist. Only for $n=1$ there exists just one value of space dimensionality, namely $D= D_{\mbox{\tiny{min}}} = 3$, for which $E_0 <0$. Another conclusion we can infer from the constraint given by equation (\ref{condition}) is that there is no bound state for hydrogen atom in space with $D=3$ whatever is the value of $n\neq 1$.  
\renewcommand{\arraystretch}{1.4}
\renewcommand{\tablename}{Table}
\begin{table}[ht]
  \caption{Predicted values for $E_{0}$ for different values of space dimensionality ($D$) and power ($n$) of the Laplacian operator.}
  \vspace*{0.2cm}
  \begin{center}
  \begin{tabular}{|c|c|c|}
    \hline
      ($D,n$)      &  $E_{0}$ (Ha)     \\  \hline
      (3,1)        & $-0.11$               \\  \hline
      (7,3)        & $-0.00041$              \\  \hline
      (8,3)        & $-6.06\times10^{-6}$      \\  \hline
      (9,3)        & $-1.52\times10^{-8}$       \\  \hline
      (10,3)       & $-1.95\times10^{-13}$       \\  \hline
      (11,3)       & $-9.92\times10^{-28}$        \\  \hline
      (11,5)       & $-1.75\times10^{-7}$          \\  \hline
      (12,5)       & $-3.23\times10^{-9}$           \\  \hline
         \vdots    & \vdots                          \\  \hline
      (18,5)       & $-5.70\times 10^{-47}$           \\  \hline
      (19,5)       & $-4.41\times 10^{-97}$            \\  \hline
  \end{tabular}
  \end{center}
  \label{e0values}
\end{table}
\renewcommand{\arraystretch}{1}

For $n=3$, for example, different bound states are found for $D=7,8,9,10$ and 11. Curiously, equation~(\ref{condition})~excludes the possibility of having $E_0 <0$ for $D=4,5$ and 6, independently of the value of $n$. Last but not least, we can see that the usual solution of the Schr\"{o}dinger equation
\begin{equation}\label{Schr-eq}
-\Delta \psi - \frac{\alpha}{r}\psi = E \psi
\end{equation}
corresponding to $D=3$ and $n=1$, represents the most bound and stable solution (See Table~\ref{e0values}) for a hydrogen atom described by equation~(\ref{wave-eq}). The specific value $E_{0}= -0.11$~Ha is the same found in Ref.~\cite{Witten} and should be compared to the well known result $-0.5$~Ha. The discrepancy found here is due to the quite modest accuracy obtained by the $1/N$ expansion when in the sequel one puts $N=3$. However, this fact does not invalidate the analysis made in this paper since the accuracy of the method is much better for high $N$ values. This is the only case where the ground state has a significant binding energy and do not show the expected features of a Rydberg-like atom, even for the first principal quantum number. For example, from Table~\ref{e0values}, we see that the value of the ground state energy found for $D=7$ and $n=3$ is equivalent to the energy of the same atom in $D=3$ and $n=1$ having a principal quantum number 35. A final comment that follows from the numbers reported in the Table is that, for a fixed value of the Laplacian operator power $n$, the ground state energy substantially decreases when the dimensionality $D$ increases.

{For completeness, although we do not believe it is the best choice, let us discuss what kind of constraint would result from another possibility: the case where, instead of making $\lambda=n$, we fix for any arbitrary $n$, $\lambda=1$. This is a more restrictive hypothesis which is frequently made in the literature. It means that in passing from $D=3$ to a space with arbitrary $D$, the potential is recognized as still being determined by the same Green equation as in $D=3$, equation~(\ref{general-poisson}) with $\lambda=1$, by just increasing the number of components of the $D$-dimensional position vector. In such case, the potential should be given by equation~(\ref{V-m1}) and, to be compatible,} equation~(\ref{E0-final}) must be replaced by
\begin{equation}\label{E0-Dn}
E_0 (D,n) = - \, \displaystyle\left[\frac{n (D/2)^{2n}}{(D/2-n)}\right]^{\frac{(D/2-n)}{(D/2-n-1)}}\times \alpha(D,1)^{-n/(D/2-n-1)}\, \times  \frac{2n-D/2}{n}
\end{equation}
Therefore, a less restrictive relationship between $n$ and $D$ is found, namely, $E_0 < 0$ if $2n+2-D > 0 \,\Rightarrow \,D < 2(n+1)$. But on the other hand, the constraint $\beta = D-2\lambda \geq 0$ reduces now to $D \geq 2$. Combining the two constraints in this case, we must have, instead of equation~(\ref{condition}), that space dimensionality should be bound by the inequality $2\leq D<2(n+1)$. So, for $n = 2$, the following dimensions are allowed: $D = 3, 4$ and 5; For $n=3$, for instance, $E_0 <0$ implies that space dimensionality should be $D=3,4,5,6$, and 7. These results (corresponding to the choice $\lambda=1$) are quite different from the case where $n=\lambda$, for which $7\leq D \leq 11$. Therefore we can conclude that, for $\lambda=1$, different negative values for $E_0$ can be found for different possible values of $D$.

\section{Concluding remarks}
\label{sec:5}
In summary, assuming that a hydrogen atom is described by equations (\ref{wave-eq}) and (\ref{pot-V-general}), with the choice $n=\lambda$, we have learned that:
\begin{enumerate}
  \item the only possibility of having a bound state in $D=3$ is with $n=1$;
  \item any other combination of $D$ and $n$ will give rise to a ground state energy which, in magnitude, will be at least $10^3$ less than the measured value 0.5~Ha;
  \item nature seems to favor 3-dimensional space and the description of hydrogen atom in terms of the usual Schr\"{o}dinger equation with the kinetic term of the Hamiltonian proportional to $\Delta$.
\end{enumerate}

{Finally, it is important to stress that we were able to shed a new light on Ehrenfest's approach trying to explain space dimensionality from a physical point of view~\cite{Ehrenfest1,Ehrenfest2}. In fact, we have not only found the possible combinations of the parameters $D$ and $n$ which correspond to a stable bound state solution of hydrogen atom but have gone further. Indeed,} to the best of our knowledge, it is the first time one has a theoretical framework in which it is possible to understand, in addition to the query of atomic stability, that the measured value for the energy of the hydrogen atom ground state, $-0.5$~Ha, is a consequence of the very particular combination: $D=3$ and $n=1$.

\section*{Acknowledgments}
This paper is dedicated to the memory of Juan Jos\'{e} Giambiagi to whom understanding space dimensionality was a scientific challenge. One of us (FS) was financed in part by the Coordena\c{c}\~{a}o de Aperfei\c{c}oamento de Pessoal de N\'{\i}vel Superior -- Brazil (CAPES), Finance Code 001.

\end{document}